\documentclass[aps, prl, nofootinbib, twocolumn]{revtex4}
\usepackage{amssymb,amsmath,microtype,url}
\usepackage{graphicx}
\usepackage[normalem]{ulem}

\usepackage{color}

\begin{document}

\title{Probing the origin of our universe through primordial gravitational waves by Ali CMB project}

\author{CAI Yi-Fu$^{1}$}
\author{ZHANG Xinmin$^{2}$}

\affiliation{$^1$CAS Key Laboratory for Researches in Galaxies and Cosmology, Department of Astronomy, University of Science and Technology of China, Chinese Academy of Sciences, Hefei, Anhui 230026, China}
\affiliation{$^2$Theoretical Physics Division, Institute of High Energy Physics, Chinese Academy of Sciences, P.O.Box 918-4, Beijing 100049, P.R.China}

\begin{abstract}
This is a research highlight invited by SCIENCE CHINA Physics, Mechanics \& Astronomy. 
\end{abstract}

\maketitle

Gravitational waves (GW), which were predicted by Einstein in 1916 based on the classical theory of General Relativity (GR), were recently detected by LIGO \cite{Abbott:2016blz}. This breakthrough is expected to initiate a novel probe of cosmology, the nature of gravity as well as fundamental physics. In general, signals of GWs can be classified into two categories, which are waves from astro-physical and cosmological sources respectively. Accordingly, a number of astronomical and cosmological experiments are under design across the world \cite{Blair:2015}. In particular, China is playing a very important role in this field by strengthening a series of fundamental scientific subjects, such as cosmic evolution, structure of matter, the origin of life, cognition science, and so on, in the 13th National Five-Year Plan\footnote{\url{http://www.gov.cn/xinwen/2016-03/17/content_5054992.htm}}. The Ali project, which aims at measuring the polarization patterns of the cosmic microwave background (CMB) radiation, was put forward in 2014 under the leadership of Xinmin Zhang's group and has become very promising in exploring primordial GWs.

Primordial GWs were produced from quantum fluctuations of our universe at the very early moments right after the Big Bang \cite{Grishchuk:1974ny}. Since the energy scale of the universe was extremely high in this epoch, the amplitudes of these tensor perturbations could be large enough to affect the physics at late times. In particular they will lead to specific imprints in the B-mode of CMB polarization. These signals, however, experience red-shifting and will damp away within the frequency band sensitive to astronomical instruments. Only the amplitudes of primordial GWs below $10^{-15}$ Hz will survive throughout the cosmic expansion, since their physical wavelengths are of the order of the observed universe today. In order to probe these signals, a major effort ought to be devoted to the design of cosmological experiments, i.e. high-precision measurements of the CMB polarization \cite{Ade:2015xua, Array:2015xqh, ZhangXM:2014}.

Inflationary cosmology suggests that the universe has experienced a short period of nearly exponential expansion in the very early times so that all unwanted primordial relics produced from the Big Bang can be diluted \cite{Guth:1981}. More importantly, during inflation the background scalar field responsible for yielding inflation gives rise to density fluctuations of quantum origin, of which the wavelengths were initially inside the Hubble radius but were stretched to super-Hubble scales by the exponential expansion of space. These primordial modes become classical and then can provide the seeds for the formation of the large-scale structure (LSS) and the CMB anisotropies \cite{Mukhanov:1992}.

In addition to primordial density fluctuations, inflation also produces primordial tensor perturbations, i.e., primordial GWs \cite{Starobinsky:1979}. These fluctuations can be described by a traceless and transverse tensor of metric perturbation $h_{ij}$ governed by a generalized Klein-Gordon equation. In Fourier space, each mode $h_k$ denoted by a fixed co-moving wave number $k$ obeys the following equation of motion:
\begin{align}
 h_k''+2\frac{a'(\tau)}{a(\tau)}h_k'+k^2h_k=0~,\nonumber
\end{align}
and its physical frequency is given by $f=k/(2\pi a)$ where $a$ is the scale factor that characterizes the size of the universe; and the prime denotes the derivative with respect to the conformal time $\tau$. From the above equation, one can read off that $h_k$ keeps oscillating when $k>a'/a$, and thus, consistently connects with the quantum state of the Bunch-Davies vacuum at the very beginning. During the inflationary expansion, $h_k$ gets squeezed on super-Hubble scales with $k<a'/a$. This causal mechanism predicts a nearly scale-invariant power spectrum for primordial GWs.

\begin{figure}
\begin{center}
\includegraphics[scale=0.45]{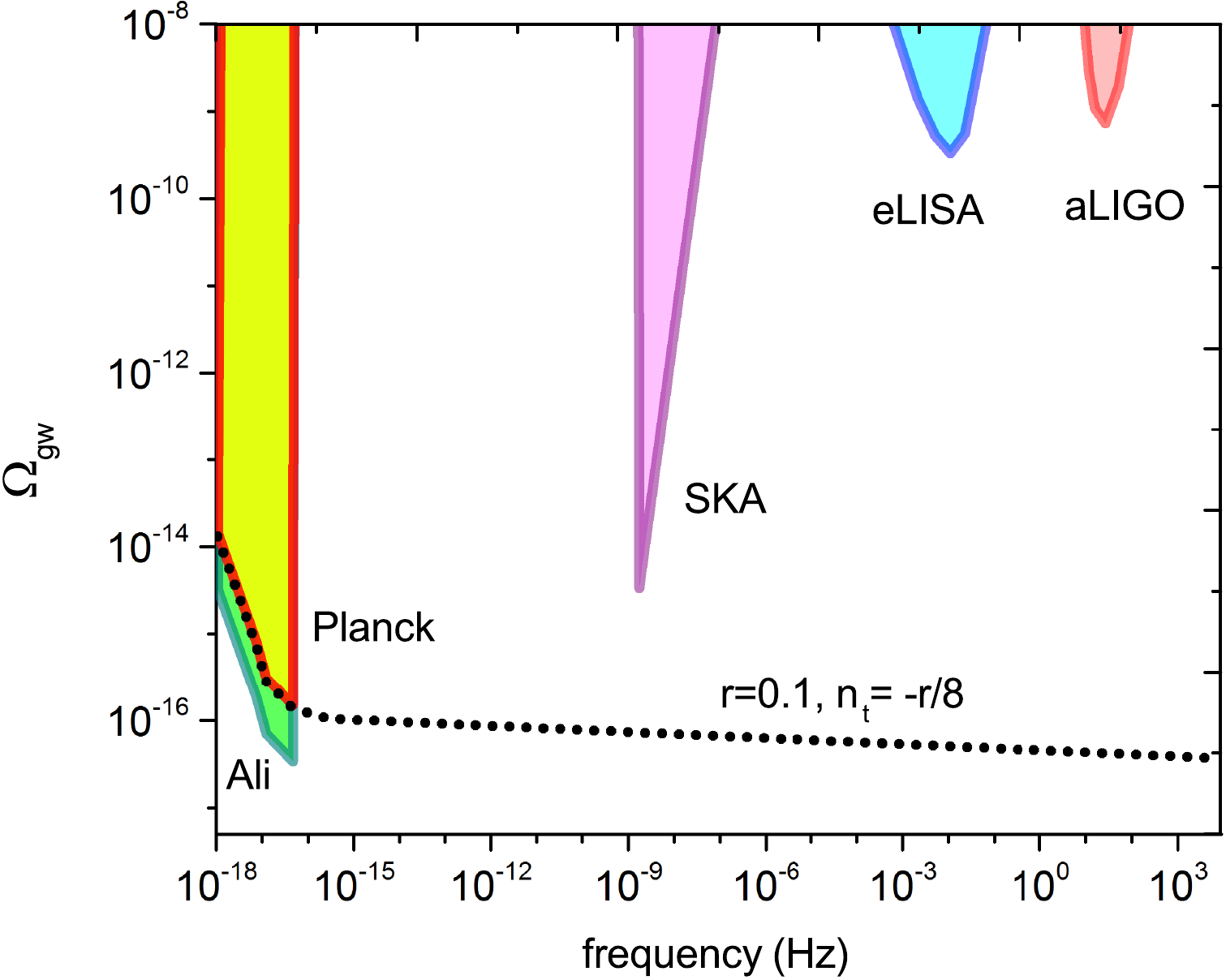}
\caption{(Color online) Theoretical curve of the energy spectrum of inflationary primordial GWs with the tensor-to-scalar ratio $r=0.1$, the slightly red spectral index $n_t=-r/8$ and its comparison with observational sensitivities of experiments including aLIGO, eLISA, SKA, Planck and Ali (under design).
}
\label{fig:gw}
\end{center}
\end{figure}

Two quantities can be applied to describe the physics of primordial GWs, namely the power spectrum $P_{gw}$ at primordial era and the energy spectrum $\Omega_{gw}$ at present. The former depicts the behavior of GWs at very early times and the latter corresponds to what can be measured in current and future observations. They are related to each other through a transfer function that characterizes the detailed information of the cosmological evolution \cite{Boyle:2005se}. One can also use the tensor-to-scalar ratio $r$, defined as the ratio of the amplitudes of power spectra for tensor and scalar modes. As an example, we consider an inflation model which yields a primordial GW spectrum with $r=0.1$. Its energy spectrum and a comparison with various experiments are provided in Figure \ref{fig:gw}.

Inflationary cosmology, however, is not the unique paradigm of the very early universe. Both the hot Big Bang and inflationary cosmology suffer from an intrinsic conceptual problem, i.e., an existence of initial cosmic singularity. According to the proof by Hawking and Penrose \cite{Penrose:1965} and later extended by Borde and Vilenkin \cite{Borde:1994}, the universe seems to begin from a space-time singularity with infinitely large energy density and temperature, at which the known physics laws would have failed. To address this issue, cosmologists are forced to search for paradigms of the very early universe beyond inflation, such as bounce cosmology \cite{CaiRev:2014}, cyclic universe \cite{LehnersRev:2008}, emergent universe \cite{Ellis:2004, Brandenberger:1989}. Within these scenarios, the big bang singularity can either be replaced by a nonsingular bounce due to quintom matter \cite{Feng:2005}, or non-canonical kinetic operators \cite{CaiBounce:2012} or a quasi-Minkowski space-time realized by string theory \cite{Brandenberger:1989}. Accordingly, the question follows that whether these cosmological paradigms can be differentiated by observations. This issue was discussed in \cite{Brandenberger:2011}, which pointed out that a careful characterization of the power spectrum of primordial B-mode polarization can be used to distinguish between the various paradigms of very early universe cosmology.

Based on cosmological perturbation theory, one can obtain a consistency relation between the spectral index $n_t$ and the tensor-to-scalar ratio $r$ within inflationary cosmology: $n_t=-r/8$ \cite{Riotto:2002}; for nonsingular bounce cosmology involving matter contraction, there is an almost exact relation: $n_t \simeq 0$ \cite{CaifT:2011}; for the emergent universe scenario realized by the string-theory thermodynamics, there is another consistency relation between the spectral index of scalar perturbations and that of tensor modes: $n_t \simeq 1-n_s$ \cite{Brandenberger:2007}; however, for a type of bounce cosmology dubbed as ekpyrosis, the power spectrum of primordial GWs is predicted to be blue with $n_t \simeq 2$ \cite{Finelli:2002}. If there was a nonsingular bounce occurred before inflation, one can also avoid the initial singularity \cite{Piao:2004} and the power spectrum of primordial GWs presents an oscillatory and damping feature at large length scales \cite{Cai:2008ed, XiaPRL:2014}. Therefore, a high-precision measurement of primordial GWs becomes very important in probing the origin of the universe.

So far, the most representative CMB experiment is the Planck project supported by the European Space Agency, which has measured the temperature anisotropies, E-mode as well as lensing B-mode at high precision. However, the Planck satellite has not yet observed any primordial B-mode but only provided an upper bound of $r<0.11$ at $2\sigma$ confidence level \cite{Ade:2015xua}. A joint analysis of the data from the BICEP2/Keck Array and Planck can improve the limit to be $r<0.07$ at $2\sigma$ \cite{Array:2015xqh, Li:2015vea}. Thus, there is a promising opportunity to explore the parameter space of primordial GWs \cite{WangMa:2014}. In particular, the Ali international collaboration
led by the Institute of High Energy Physics at Chinese Academy of Sciences
will be the first ground-based CMB experiment in the northern hemisphere of the Earth and is expected to improve the upper bound at the level of 1\%.

We end the article by highlighting the scientific importance of the Ali CMB project under design. Over the past couple of decades, modern cosmology has made a number of successful achievements. However, physicists still do not understand the origin of the universe. The standard paradigm of inflationary Big Bang cosmology can interpret the observational data precisely but suffers from the initial singularity problem. Non-singular cosmology can avert this puzzle. Both Inflation and non-singular models predict the existence of primordial GWs, but with different patterns. By detecting and measuring them in CMB polarization experiments, we have the chance to probe the origin of the universe. This is one major goal of the Ali project under design in Tibet of China.
The high-precision B-mode measurement will enable us to extract detailed information of primordial GWs, and hence we may grasp the chance to differentiate various scenarios of very early universe, and then our knowledge about our universe will be extended to the regime close to that of quantum gravity, a regime of science which has never before been explored.

\textit{Acknowledgements.--}
Authors thank R. Brandenberger, J. Chen, X. Chen, Z. Fan, Y. Gong, Z. Guo, Q. Huang, H. Li, M. Li, S. Li, H. Liu, T. Qiu, M. Su, Y. Wan, D. Wang, K. Wang, Y. Wang, L. Zhang, P. Zhang, X. Zhang, W. Zhao for valuable comments. CYF is supported in part by the Chinese National Youth Thousand Talents Program, by the USTC start-up funding (KY2030000049) and by the NSFC (Grant No. 11421303). XZ is supported in part by the NSFC (Grant Nos. 11121092, 11033005, 11375220) and by the CAS pilot-B program.

\end{document}